\documentclass{aastex631}
\usepackage{subscript}

\usepackage{xcolor}
\usepackage{soul}
\graphicspath{{./}{figures/}}

\newcommand{\aco}{\alpha_{\rm{CO}}}

\newcommand{\beq}{\begin{equation}}
\newcommand{\eeq}{\end{equation}}

\newcommand{\Msun}{\rm M_\odot}
\newcommand{\kmps}{km~s$^{-1}$}

\newcommand{\Mmol}{{\rm M_{mol}}}
\newcommand{\acou}{$\Msun$~(K~\kmps~pc$^2)^{-1}$}

\newcommand{\hi}{H{\sc i}}

\newcommand{\cii}{[C{\sc ii}]\,158$\mu$m}

\begin{document}

\title{Jansky Very Large Array detections of CO(1--0) emission in H{\sc i}-absorption-selected galaxies at $z \gtrsim 2$}

\correspondingauthor{Balpreet Kaur}
\email{bkaur@ncra.tifr.res.in}

\author{B. Kaur} 
\affiliation{National Centre for Radio Astrophysics, Tata Institute of Fundamental Research, 
Pune University, Pune 411007, India}

\author{N. Kanekar} 
\affiliation{National Centre for Radio Astrophysics, Tata Institute of Fundamental Research, Pune University, Pune 411007, India}

\author{M. Rafelski}
\affiliation{Space Telescope Science Institute, 3700 San Martin Drive, Baltimore, MD 21218, USA}
\affiliation{Department of Physics \& Astronomy, Johns Hopkins University, Baltimore, MD 21218, USA}

\author{M. Neeleman}
\affiliation{Max-Planck-Institut für Astronomie, Königstuhl 17, D-69117, Heidelberg, Germany}

\author{J. X. Prochaska}
\affiliation{Department of Astronomy \& Astrophysics, UCO/Lick Observatory, University of California, 1156 High Street, Santa Cruz, CA 95064, USA}
\affiliation{Kavli Institute for the Physics and Mathematics of the Universe (Kavli IPMU), 5-1-5 Kashiwanoha, Kashiwa, 277-8583, Japan}

\author{M. Revalski}
\affiliation{Space Telescope Science Institute, 3700 San Martin Drive, Baltimore, MD 21218, USA}

\begin{abstract}
We report a Karl G. Jansky Very Large Array search for redshifted CO(1--0) emission from three  H{\sc i}-absorption-selected galaxies at $z \approx 2$, identified earlier in their CO(3--2) or CO(4--3) emission. We detect CO(1--0) emission from DLA~B1228-113 at $z\approx2.1933$ and DLA~J0918+1636 at $z\approx2.5848$; these are the first detections of CO(1--0) emission in high-$z$ H{\sc i}-selected galaxies. We obtain high molecular gas masses, $\rm M_{mol}\approx10^{11}\times(\alpha_{\rm CO}/4.36)\ M_\odot$, for the two objects with CO(1--0) detections, which are a factor of $\approx1.5-2$ lower than earlier estimates. We determine the excitation of the mid$-J$ CO rotational levels relative to the $J=1$ level, r$_{ J1}$, in H{\sc i}-selected galaxies for the first time, obtaining r$_{\rm 31}=1.00\pm0.20$ and r$_{\rm 41}=1.03\pm0.23$ for DLA~J0918+1636, and r$_{\rm 31}=0.86\pm0.21$ for DLA~B1228-113. These values are consistent with thermal excitation of the $J=3,4$ levels. The excitation of the $J=3$ level in the \hi-selected galaxies is similar to that seen in massive main-sequence and sub-mm galaxies at $z\gtrsim2$, but higher than that in main-sequence galaxies at $z\approx1.5$; the higher excitation of the galaxies at $z\gtrsim2$ is likely to be due to their higher star-formation rate (SFR) surface density. We use Hubble Space Telescope Wide Field Camera~3 imaging to detect the rest-frame near-ultraviolet emission of DLA~B1228-113, obtaining an NUV SFR of $4.44\pm0.47\ \Msun$~yr$^{-1}$,  significantly lower than that obtained from the total infrared luminosity, indicating significant dust extinction in the $z\approx2.1933$ galaxy.
\end{abstract}


\section{Introduction} \label{sec:intro}

The highest \hi\ column density absorption systems in QSO spectra, the damped Lyman-$\alpha$ absorbers (DLAs), arise from gas associated with high-redshift galaxies \citep[e.g.][]{Wolfe05}. Such absorbers provide a route to identifying high-$z$  galaxies without the luminosity bias that afflicts galaxy samples selected directly via their emission. Understanding the nature of these \hi-absorption-selected galaxies is critical for an unbiased understanding of galaxy evolution. Unfortunately, the proximity of the faint foreground galaxy to the bright background QSO has meant that, despite many studies, it has been difficult to even identify, let alone characterize, the high-$z$, \hi-selected galaxies via optical imaging and spectroscopy \citep[e.g.][]{Fumagalli15,Krogager17}.

Recently, the Atacama Large Millimeter/submillimeter Array (ALMA) and the NOrthern Extended Millimeter Array (NOEMA) has opened new windows on \hi-selected galaxies via their redshifted \cii\ and CO emission. Besides the characterization of \hi-selected galaxies at intermediate redshifts, $z \approx 0.6$ \citep[e.g.][]{Moller18,Kanekar18,Peroux19,Klitsch19}, this has resulted in the identification of more than a dozen such galaxies at $z \approx 2$ \citep{Neeleman18,Fynbo18,Kanekar20,Kaur22} and $z \approx 4$ \citep{Neeleman17,Neeleman19}. Perhaps the most remarkable results from these studies are the large impact parameters to the QSO sightline  of the galaxies at $z \approx 4$ \citep{Neeleman17,Neeleman19}, the high molecular gas masses of the galaxies at $z \approx 2$ \citep{Kanekar20}, and the identification of a cold, dusty, rotating disk at $z \approx 4.26$ \citep{Neeleman20}. Multi-wavelength observations are now under way to obtain a detailed understanding of these galaxies, by characterizing their stellar and gas properties,  star-formation activity, etc \citep[e.g.][]{Prochaska19,Kaur21,Klitsch22}.

For the $z \approx 2$ \hi-selected galaxies, the high inferred molecular gas masses, $\Mmol \approx (1.3 - 20.7) \times 10^{10} \ \Msun$ \citep{Kanekar20}, were obtained from the CO(3--2) or CO(4--3) mid-$J$ rotational transitions. These estimates are based on two critical assumptions, sub-thermal excitation of the CO ${J}=3$ or ${J}=4$ level relative to the ${J}=1$ level, and the value of the CO-to-H$_2$ conversion factor, $\aco$ \citep[e.g.][]{Tacconi20}. Direct observations of the CO(1--0) line are critical to remove the first of these assumptions, and obtain more accurate molecular gas mass estimates. Combining the CO(1--0) line luminosity with the luminosity in the higher-$J$ CO lines also allows one to determine the CO excitation and thus probe physical conditions in the molecular gas \citep[e.g.][]{Carilli13}. Indeed, studies in the local Universe have shown that the CO excitation is related to the value of $\aco$: high-excitation galaxies like ultra-luminous infrared galaxies (ULIRGs) and QSO hosts have lower $\aco$ values [$\aco \approx 1.0 \ \Msun$~(K~\kmps~pc$^2$)$^{-1}$] than low-excitation objects like the Milky Way and nearby disk galaxies [$\aco \approx 4.3 \ \Msun$~(K~\kmps~pc$^2$)$^{-1}$; e.g.  \citealp{Bolatto13}].

At high redshifts, CO(1--0) studies have so far only been carried out in bright, emission-selected galaxies, mostly QSO hosts, sub-mm galaxies, lensed galaxies, and massive main-sequence galaxies \citep[e.g.][]{Aravena14,Bolatto15,Sharon16,Riechers20}. 
In this {\it Letter}, we report the first detections of redshifted CO(1--0) emission from \hi-selected galaxies at $z \approx 2$, obtained with the Karl G. Jansky Very Large Array (JVLA)\footnote{We assume a flat $\Lambda$-cold dark matter cosmology, with $\Omega_{\Lambda} = 0.685$, $\Omega_{m} = 0.315$, and H$_0 = 67.4$~\kmps~Mpc$^{-1}$ \citep{Planck2020}.}. We also report Hubble Space Telescope (HST) Wide Field Camera~3 (WFC3) imaging of the rest-frame near-ultraviolet continuum emission for one of the \hi-selected galaxies, which shows that the object is a dusty galaxy.

\begin{table*}
\centering
\caption{JVLA Ka-band observations and results. The columns are (1)~the galaxy name, (2)~the QSO redshift, (3)~the absorption redshift, $z_{\rm abs}$, (4)~the galaxy redshift, $z_{\rm gal}$, from the higher-$J$ CO lines, (5)~the redshifted CO(1--0) line frequency, in GHz, (6)~the synthesized beam, in $'' \times ''$, (7)~the RMS noise, in $\mu$Jy/Beam, at a velocity resolution of 100~\kmps, (8)~the integrated CO line flux density, in Jy~\kmps, (9)~the CO(1--0) line luminosity, $L{\rm '_{CO(1-0)}}$, and (10)~the molecular gas mass, $\Mmol$, assuming $\aco = 4.36$~\acou. For the CO(1--0) non-detection in B0551-366, the last three columns list the $3\sigma$ upper limits on $\int S_\mathrm{CO} \ dV$, $L{\rm '_{CO(1-0)}}$, and $\Mmol$, for an assumed line FWHM of 300~\kmps\ \citep{Kanekar20}.
\label{tab:obs}}
\vspace{0.2cm}    
\begin{tabular}{lcccccccccc}
\hline
\hline
DLA & $z_{\rm QSO}$ & $z_{\rm abs}$& $z_{\rm gal}$  & $\nu_{\rm obs}$  & Beam & $\rm RMS_{CO}$ &  $\int S_\mathrm{CO} \ dV$  &  $L\rm'_{CO(1-0)}$ & $\Mmol$ \\
& &  & & GHz  & $''\times''$  & $\mu$Jy/Beam  & Jy~\kmps  & $10^{10}$~K~\kmps~pc$^2$ & $10^{10} \ \Msun$ \\

\hline 
B0551-366  & 2.317 & 1.9622 & 1.9615 & 38.92 & $5.8 \times 1.6$ & 99  & $<0.057$         & $<1.14$         & $< 5.0$         \\
B1228-113  & 3.528 & 2.1929 & 2.1933 & 36.10 & $3.1 \times 2.0$ & 94 & $0.094 \pm 0.023$ & $2.29 \pm 0.56$ & $10.0 \pm 2.4$  \\
J0918+1636 & 3.096 & 2.5832 & 2.5848 & 32.15 & $2.8 \times 2.3$ & 57 & $0.082 \pm 0.016$ & $2.65 \pm 0.51$ & $11.5 \pm 2.2$  \\

\hline

\end{tabular}
\vskip 0.1in

\end{table*}

\section{Observations, Data Analysis, and Results}
\label{sec:obs}

\subsection{JVLA observations and data analysis}
\label{subsec:jvla}

We used the JVLA Ka-band receivers in the D-array to search for redshifted CO(1--0) emission from the \hi-selected galaxies at $z \approx 1.9615$ towards QSO~B0551-366, $z \approx 2.1933$ towards PKS~B1228-113, and $z \approx 2.5848$ towards QSO~J0918+1636 in 2019 November and December  (Proposal ID: VLA/19B-271, PI: N.~Kanekar). The three galaxies (hereafter, DLA~B0551-366, DLA~B1228-113 and DLA~J0918+1636) had been earlier detected with ALMA in their CO(4--3) or CO(3--2) emission \citep{Neeleman18,Fynbo18,Kanekar20}. 

The JVLA observations used the WIDAR correlator in 8-bit mode, with two 1-GHz intermediate frequency (IF) bands, one covering the redshifted CO line frequency for the target galaxy, and two polarizations. Each IF band was divided into eight 128-MHz digital sub-bands, with each sub-band further divided into 512 and 128 channels, for line and off-line sub-bands, respectively. This yielded a velocity coverage of $\approx 800 - 955$~\kmps\ for the sub-band covering the redshifted CO line frequency, albeit with lower sensitivity at the sub-band edges, and a raw velocity resolution of $\approx 2$~\kmps. Observations of the targets were interleaved with runs on nearby phase calibrators every 5 minutes; in addition, a standard flux calibrator was observed once in each session. The on-source times were $2.1-5.8$~hrs for the three galaxies.

All data were analysed in the Astronomical Image Processing System package \citep[``classic'' {\sc aips};][]{Greisen03}, using standard procedures. After identifying and editing out malfunctioning antennas and any data affected by systematic errors, we used the calibrator data to determine the antenna-based complex gains and bandpass shapes. The gains and bandpasses were applied to the data for each target, and the calibrated target visibilities then split out into a new data set. For each target, the central channels of each sub-band of this data set were averaged together, and the resulting data sets were imaged to search for any continuum emission in each field. Although continuum emission was detected from two QSOs, only PKS~B1228-113 was found to have a sufficient flux density for self-calibration. The self-calibration for PKS~B1228-113 followed a standard iterative procedure, with multiple rounds of imaging and phase-only self-calibration, followed by  amplitude-and-phase self-calibration and data editing, until neither the image nor the residual visibilities showed any improvement on further self-calibration. The improved complex gains for DLA~B1228-113 were applied to the multi-channel data to produce the final calibrated data set for this field.

Next, for each target, the task {\sc uvsub} was used to subtract out all continuum emission from the calibrated visibilities. The residual multi-channel visibility data sets were then 
imaged in the Common Astronomy Software Applications package \citep[{\sc casa} version~5.6;][]{CASA}, using natural weighting in the barycentric frame, to produce spectral cubes at velocity resolutions of $50-100$~\kmps. The full-width-at-half-maximum (FWHM) of the JVLA synthesized beams is $\approx 1.6''-5.8''$ for the three targets, larger than the size of the ALMA CO emission \citep{Kanekar20}; we hence do not expect any of the CO(1--0) emission to be resolved out in the JVLA images. For each galaxy, a CO(1--0) emission spectrum was extracted by taking a cut through its cube at the location of the ALMA CO emission. Finally, in the case of DLA~B1228-113 and DLA~J0918+1636, which showed detections of CO(1--0) emission at $> 4\sigma$ significance, we also made images of the CO emission, at velocity resolutions matched to the line FWHMs. The observational details and results are summarized in Table~\ref{tab:obs}.

\subsection{HST observations and data analysis}
\label{ref:hst}

The HST WFC3 observations of DLA~B1228-113 were carried out in late 2019 (PID: 15882; PI: Kanekar), using the F105W filter to cover the rest-frame NUV stellar continuum from the $z \approx 2.1933$ galaxy. A single orbit was obtained, using a WIDE-7 dither pattern increased by a factor of 3 over the pattern described in the Instrument Science Report \citep[ISR 2016-14;][]{Anderson:2016}, in order to obtain cleaner images by dithering over the size of the IR blobs. 

 The data were calibrated using the new IR filter-dependent sky flats \citep[WFC3 ISR 2021-01; ][]{Mack21}. Image mosaics were made using AstroDrizzle \citep{hack20}, drizzling to a scale of $0\farcs06$~pixel$^{-1}$. TweakReg was used to astrometrically align the image to the GAIA DR2 catalog \citep{Gaia18}, yielding an absolute astrometric uncertainty of $\approx 0\farcs01$. The effective angular resolution of the HST WFC3 image is $\approx 0.3''$, based on a Gaussian fit to the point spread function (PSF) of the quasar.

\subsection{Results}
\begin{figure*}
\centering
\includegraphics [scale=0.10]{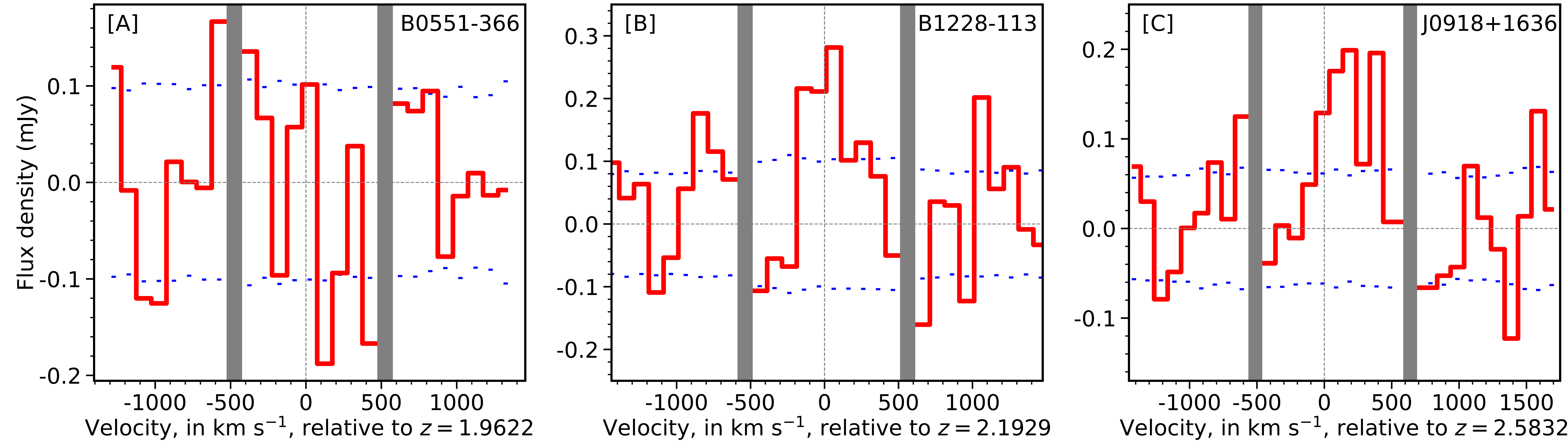}
\caption{JVLA CO(1--0) spectra from [A]~DLA~B0551-366 at $z = 1.9615$, [B]~DLA~B1228-113 at $z = 2.1933$, and [C]~DLA~J0918+1636 at $z = 2.5848$. 
The vertical grey bands indicate the edges of the digital sub-bands of the WIDAR correlator, which have significantly lower sensitivity.
The blue dashed curves in each panel indicate the $\pm 1 \sigma$ error at each velocity channel. 
\label{fig:spectra}}
\end{figure*}

\begin{figure*}
\centering
\includegraphics [scale = 0.17]{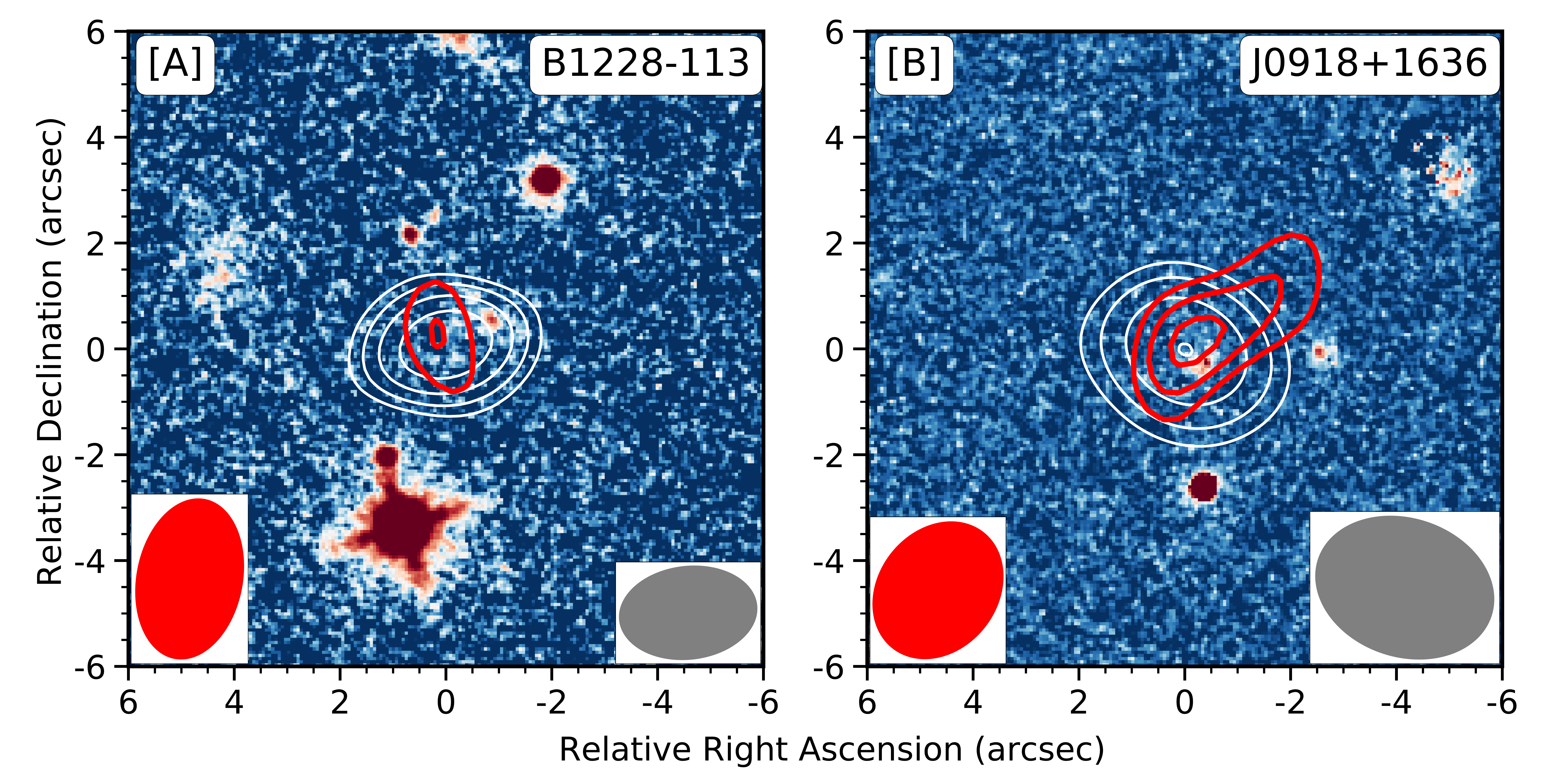}
\caption{{JVLA CO(1--0) integrated emission (red contours) overlaid on the ALMA CO(3--2) emission (white contours) and the HST WFC3 F105W image (in colour) for [A]~DLA~B1228-113 at $z = 2.1933$ and [B]~J0918+1636 at $z=2.5848$ respectively. The left and right insets in each panel show, respectively, the JVLA (red) and ALMA (grey) synthesized beams. The CO(1--0) contours are at $(-3.0, 3.0, 4.0, 5.0) \times \sigma$ significance (note that there are no  negative contours with $\geq 3\sigma$ significance), while the outermost CO(3--2) contour is at $6\sigma$ significance, with successive later contours increasing by factors of $\sqrt{2}$.}
\label{fig:images}}
\end{figure*}

Our JVLA observations yielded detections of CO(1--0) emission at $> 4\sigma$ significance from the \hi-selected galaxies at $z \approx 2.1933$ towards PKS~B1228-113 and $z \approx 2.5848$ towards QSO~J0918+1636. We also obtained an upper limit to the CO(1--0) line luminosity for the $z \approx 1.9615$ galaxy towards QSO~B0551-366. Fig.~\ref{fig:spectra} shows the JVLA CO(1--0) spectra of the three \hi-selected galaxies, while Fig.~\ref{fig:images} shows the CO(1--0) images of DLA~B1228-113 and DLA~J0918+1636 (in red contours), overlaid on the ALMA CO(3--2) images (white contours), and the HST WFC3 images \citep[in colour; ][this work]{Fynbo18} of the two fields. For DLA~B1228-113 and DLA~J0918+1636, the detected CO(1--0) emission agrees in both position and velocity with the ALMA CO(3--2) emission. The measured velocity-integrated CO(1--0) line flux densities and inferred CO line luminosities for the three galaxies are listed in Table~\ref{tab:obs}. Here, the $3\sigma$ upper limits for DLA~B0551-366 assume that the CO(1--0) emission has an FWHM of 300~\kmps, equal to that of the CO(4--3) emission \citep{Kanekar20}. The CO(1--0) line luminosities yield molecular gas masses of $(10.0 \pm 2.4) \times (\aco/4.36) \times 10^{10} \ \Msun$ (DLA~B1228-113), $(11.5 \pm 2.2) \times (\aco/4.36) \times 10^{10} \ \Msun$ (DLA~J0918+1636) and $< 5.0 \times (\aco/4.36) \times 10^{10} \ \Msun$ ($3\sigma$ limit; DLA~B0551-366), where we have assumed $\aco = 4.36$~\acou\ \citep{Bolatto13,Tacconi20}.

Both DLA~J0918+1636 and DLA~B1228-113 are detected in HST rest-frame NUV imaging. The HST images of DLA~J0918+1636 are presented and described in \citet{Fynbo18}, and yield a half-light radius of $0\farcs30$ (Fynbo, private communication). In the case of DLA B1228-113, the rest-frame NUV emission is detected in our HST WFC3 F105W image (see Fig.~\ref{fig:images}[A]), at RA=12h30m55.44s, Dec.=-11$^\circ$39$\arcmin$05.87$\arcsec$, at an offset of $\approx 0\farcs$9 ($\approx7.7$~kpc) from the CO(3--2) emission. We emphasize that no rest-frame NUV emission is detected at the center of the CO emission (see the discussion in Section~\ref{sec:discussion}).

We obtained photometry of DLA~B1228-113 using Source Extractor~v.2.19.5 \citep{Bertin96}, measuring the total flux using flux$_{\rm auto}$, which gives the flux within an elliptical aperture with the Kron radius \citep{Kron80}. This yielded an AB magnitude of $\rm m_{AB}  \approx 24.72 \pm 0.12$, a half-light radius of $0\farcs33$, and a Kron radius of $0\farcs46$. To infer the SFR of the galaxy, we assume that it has a Chabrier initial mass function \citep{Chabrier03} and a flat spectrum in $L_\nu$ between rest-frame wavelengths of $\approx 3280$\AA\ and $2300$\AA\ \citep{Kennicutt98}. Applying the local relation between rest-frame NUV $2300$\AA\ luminosity and SFR \citep{Kennicutt12} then yields an SFR of $4.44 \pm 0.47 \ \Msun$~yr$^{-1}$. 

Finally, we combined the rest-frame NUV half-light radii of DLA~B1228-113 (r$_{1/2} = 0\farcs33$) and DLA~J0918+1636 (r$_{1/2} = 0\farcs30$) with the galaxy SFRs (87~$\rm M_\odot \ yr^{-1}$ and 229~$\rm M_\odot \ yr^{-1}$, obtained from SED fits; \citealp{Klitsch22}), to obtain SFR surface densities of 1.8~$\rm M_\odot \ yr^{-1} \ kpc^2$ (DLA~B1228-113) and 6.2~$\rm M_\odot \ yr^{-1} \ kpc^2$ (DLA~J0918+1636).

\section{Discussion}
\label{sec:discussion}

\begin{table*}
\centering
\caption{The CO excitation of the three \hi-selected galaxies at $z \approx 2$. The columns are (1)~the galaxy name, (2)~the galaxy redshift, $z_{\rm gal}$, (3)~the ratio r$_{31}$, (4)~the ratio r$_{41}$, (5)~the ratio r$_{51}$, (6)~the ratio r$_{61}$, and (7)~references for the higher-$J$ CO studies. 
\label{tab:results}}
\vspace{0.2cm}    
\begin{tabular}{lcccccc}
\hline
\hline
DLA & $z_{\rm gal}$ &r$_{31}$ & r$_{41}$ & r$_{51}$ & r$_{61}$ & Refs. \\
\hline 

B0551-366 & 1.9615  &$ -$ &$>0.48$ & $>0.24$ & $>0.07$    & 1,2 \\
B1228-113 & 2.1933  & $0.86 \pm 0.21$ &  $-$    &$-$& $0.18\pm0.05$ & 1--3 \\
J0918+1636 & 2.5848  &  $1.00 \pm 0.20$ & $1.03 \pm 0.23$   & $0.39\pm0.09$ & $-$ & 1,2,4\\
\hline
\end{tabular}
\vskip 0.1in
(1)~\citet{Kanekar20}, (2)~\citet{Klitsch22}, (3)~\citet{Neeleman18}, (4)~\citet{Fynbo18}.
\end{table*}

\subsection{Molecular gas masses}

We have obtained the first detections of CO(1--0) emission in high-redshift \hi-selected galaxies. The measured CO(1--0) line luminosities yield direct estimates of the molecular gas mass, without the need for assumptions regarding the CO excitation. We continue to obtain high molecular gas masses in DLA~B1228-113  and DLA~J0918+1636, albeit a factor of $\approx 1.5$ lower than the estimates from the CO(3--2) line \citep{Neeleman18,Fynbo18,Kanekar20}. For DLA~B0551-366, we have placed an upper limit on its CO(1--0) line luminosity, and thus, on its molecular gas mass, $\Mmol < 5.0 \times (\aco/4.36) \times 10^{10} \ \Msun$. This upper limit is again slightly lower than (but, given the uncertainties, consistent with) the estimate $\Mmol = (5.67 \pm 0.68) \times (\aco/4.36) \times 10^{10} \ \Msun$, obtained by \citet{Kanekar20} from the CO(4--3) line.

Our molecular gas masses, inferred from the CO(1--0) line, are thus systematically lower, by a factor of $\approx 1.5$, than the values obtained from the mid-$J$ rotational lines \citep{Kanekar20}. This indicates that the three \hi-selected galaxies do not have substantial amounts of cold, low-excitation molecular gas \citep[e.g.][]{Papadopoulos01}. The lower gas mass estimates obtained here are because the earlier studies assumed sub-thermal excitation of the mid-$J$ CO rotational levels to estimate the CO(1--0) line luminosity, and thence the molecular gas mass. 

The main remaining uncertainty in the molecular gas mass estimates for our \hi-selected galaxies is the value of the CO-to-H$_2$ conversion factor, $\aco$. Estimates of the stellar mass and SFR from fits to the broad-band spectral energy distribution \citep[SED; ][]{Klitsch22} find that both DLA~B1228-113 and DLA~J0918+1636 lie slightly ($\approx 0.5-0.6$~dex) above the galaxy main sequence at their redshifts, but within the spread of the main sequence \citep[e.g.][]{Whitaker12,Schreiber15}, especially given the large errors in the current SFR estimates. Thus, at present, there is no significant evidence that DLA~B1228-113 or DLA~J0918+1636 are starburst galaxies, that might have $\aco \ll 4$~\acou\ \citep[e.g.][]{Magdis12}. Further, the galaxy metallicities inferred from the stellar mass -- metallicity relation at $z \approx 2$ are consistent with solar metallicity \citep{Klitsch22}. We will hence assume $\aco \approx 4.36$~\acou\ for these systems \citep[e.g.][]{Bolatto13,Tacconi20}, applicable to high-$z$ main-sequence galaxies with solar metallicity.

Combining the SFR estimates of \citet{Klitsch22} with our molecular gas mass estimates, we obtain a molecular gas depletion timescale $\tau_{\rm dep, H_2}$ of $\approx 1.1$~Gyr (DLA~B1228-113) and $\approx 0.5$~Gyr (DLA~J0918+1636). These are  factors of $\approx 2-4$ higher than the expected depletion times ($\approx 0.25$~Gyr) from the scaling relation of \citet{Tacconi20}. 

\subsection{The CO spectral line energy distribution (SLED) in \hi-selected galaxies}

Our JVLA CO(1--0) measurements allow us, for the first time, to determine the excitation of the mid-$J$ rotational levels relative to the $J=1$ level in \hi-selected galaxies. Table~\ref{tab:results} lists the values of $r_{J1} \equiv L'_{{\rm CO}[J \rightarrow (J-1)]} / L'_{\rm CO(1-0)}$, for $J = 3-6$, combining our measured CO(1--0) line luminosities with the luminosities of the higher-$J$ lines \citep{Kanekar20, Klitsch22}.

\begin{figure*}
\centering
\includegraphics [scale=0.21]{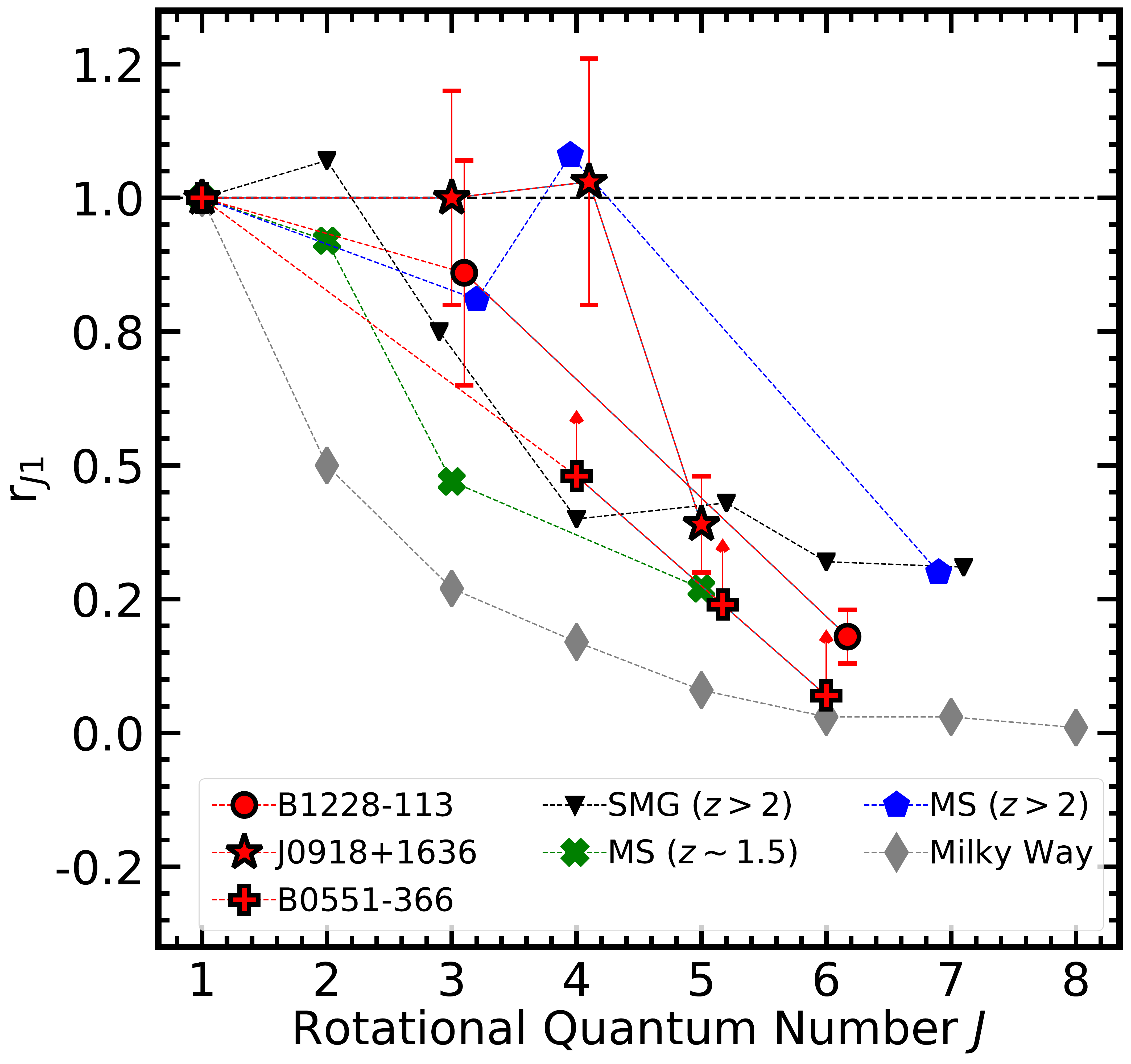}
\caption{The ratio of the CO($J~\rightarrow J-1$) line luminosity to the  CO(1--0) line luminosity, plotted as a function of the upper-state rotational quantum number, $J$. The dashed black line indicates the value of unity for thermal excitation. The red symbols represent the \hi-selected galaxies at $z \sim 2$ \citep[][this work]{Kanekar20, Klitsch22}. For comparison, we have plotted the median value of the above ratios for main-sequence galaxies at $z \approx 1.5$ \citep[green crosses; ][]{Daddi15}, main-sequence galaxies at $z > 2$ \citep[blue pentagons; ][]{Riechers10,Brisbin19, Boogaard20, Brocal21}, and submillimetre galaxies at $z > 2$ \citep[inverted black triangles; ][]{Sharon16, Calistrorivera18, Birkin21}. The grey diamonds indicate the CO SLED of the inner disk of the Milky Way \citep[][]{Fixsen99}. \label{fig:cosled}}
\end{figure*}

\begin{figure*}
\centering
\includegraphics [scale=0.21]{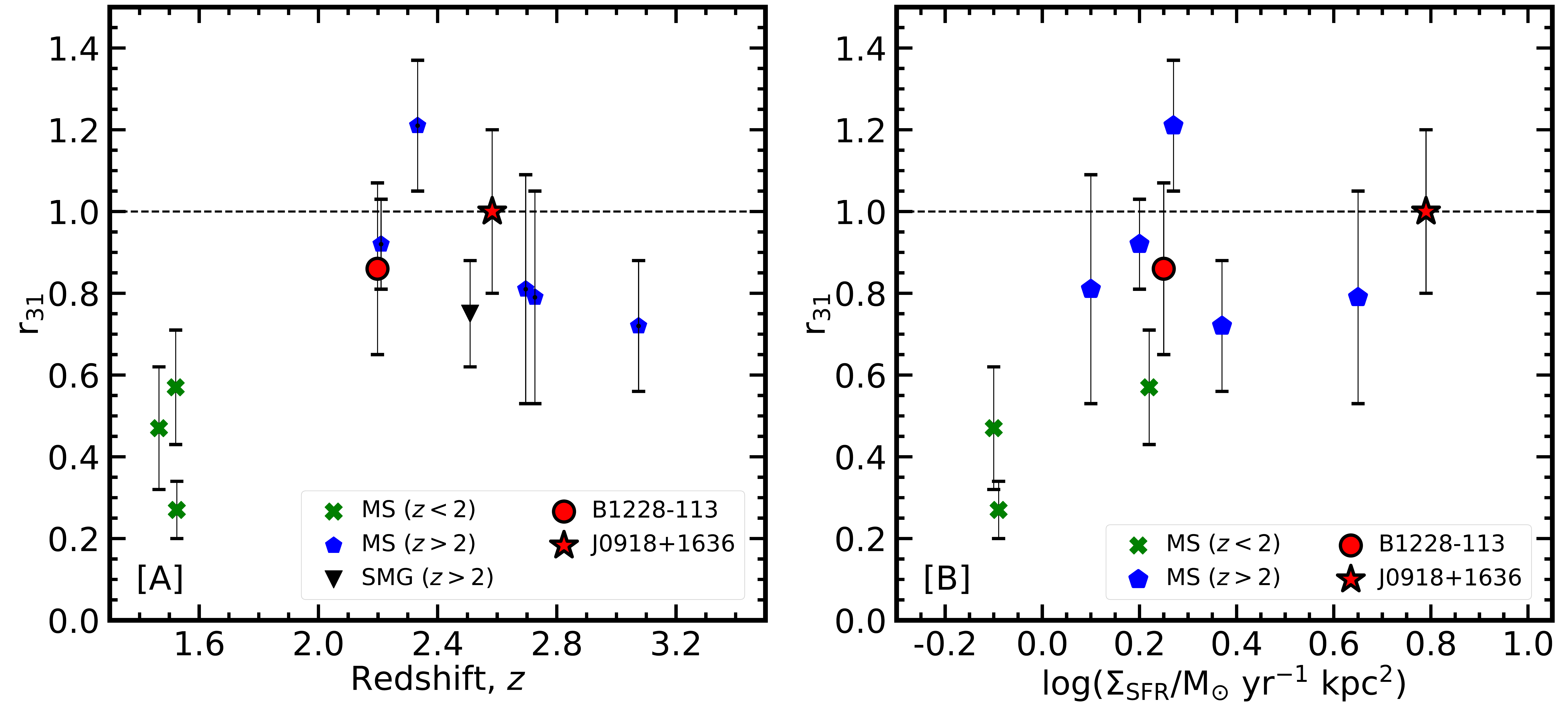}
\caption{The ratio of the CO(3--2) line luminosity to the CO(1--0) line luminosity (r$_{31}$), plotted as a function of [A]~redshift and [B]~SFR surface density, for the \hi-selected galaxies of this work (red symbols), BzK galaxies at $z \approx 1.5$ \citep[green crosses; ][]{Daddi15}, and main-sequence galaxies at $z > 2$ \citep[blue pentagons; ][]{Riechers10,Brisbin19, Boogaard20, Brocal21}. The left panel also shows the median r$_{31}$ value for submillimetre galaxies at $z > 2$ \citep[inverted black triangles; ][]{Sharon16}. The dashed horizontal line in both panels indicate r$_{31} = 1$, for thermal excitation.
\label{fig:co31}}
\end{figure*}

Studies of high-redshift galaxies typically assume sub-thermal excitation of the mid-$J$ rotational levels, with r$_{31} \approx 0.55$ and r$_{41} \approx 0.42$ \citep[e.g.][]{Tacconi20}. Remarkably, we find that DLA~J0918+1636 has values of r$_{31} = 1.00 \pm 0.20$ and r$_{41} = 1.03\pm0.23$, consistent with thermal excitation of the $J= 3, 4$ levels. We note that \citet{Klitsch22} had earlier measured r$_{43} = 1.03$ in DLA~J0918+1636, suggesting that the $J=4$ and $J=3$ levels are likely to show thermal excitation. Similarly, the value of r$_{31} = 0.86 \pm 0.21$ in DLA~B1228-113 is consistent with thermal excitation, while the CO(1--0) non-detection in DLA~B0551-366 yields the lower limit r$_{41} > 0.48$, higher than the canonical value of 0.42 \citep{Tacconi20}. We thus find direct evidence that the mid-$J$ rotational levels of massive \hi-selected galaxies at $z \gtrsim 2$ show relatively high excitation \citep[see also][]{Klitsch22}.

Fig.~\ref{fig:cosled} plots the r$_{J1}$ values of the three \hi-selected galaxies (red symbols) against the upper rotational level quantum number, $J$. The dashed horizontal line at r$_{J1} = 1$ indicates thermal excitation of the rotational levels. The figure also includes data for the inner disk of the Milky Way \citep[grey diamonds; ][]{Fixsen99}, three main-sequence (BzK) galaxies at $z \approx 1.5$ \citep[green crosses; ][]{Aravena14,Daddi15}, five main-sequence galaxies at $z > 2$ \citep[blue pentagons; ][]{Riechers10,Bolatto15,Brisbin19, Boogaard20,Riechers20,Brocal21}, and a large sample of sub-mm galaxies (SMGs) at $z > 2$ \citep[inverted black triangles; ][]{Birkin21}. For the high-$z$ main-sequence galaxies and SMGs, the figure shows the median values of r$_{J1}$ of objects with measurements of both the CO(1--0) line and higher-$J$ CO lines. In the case of the SMGs, we have used the median CO SLED of the large SMG sample of \citet{Birkin21}, who assumed r$_{21} = 0.9$ due to the paucity of CO(1--0) measurements in their sample. We bypass this assumption by normalizing their r$_{J1}$ values  to r$_{31} = 0.75$, the median measured r$_{\rm 31}$ value obtained for 18 SMGs at $z \approx 2.2-3.1$ \citep{Sharon16, Calistrorivera18}. 

Fig.~\ref{fig:cosled} shows that the three \hi-selected galaxies have CO SLEDs consistent with those of massive main-sequence galaxies at $z > 2$. The excitation of the mid-$J$ rotational levels  of the \hi-selected galaxies is clearly higher than that of both the inner disk of the Milky Way and main-sequence galaxies at $z \approx 1.5$. Further, the roughly thermal excitation of the $J = 3, 4$ levels in DLA~J0918+1636 (and of the $J=3$ level in DLA~B1228-113) is consistent with the excitation of the same levels in the main-sequence galaxy BX610 \citep[]{Bolatto15,Brisbin19}. It is interesting that the excitation in DLA~J0918+1636 drops sharply in the $J=5$ rotational level, with r$_{51} \approx 0.39$, well below thermal.
 
 Fig.~\ref{fig:co31}[A] plots r$_{\rm 31}$ versus redshift for DLA~B1228-113, DLA~J0918+1636, and the above samples of SMGs at $z > 2$, and main-sequence galaxies at $z \approx 1.5$ and $z > 2$. Here, the individual r$_{31}$ values are plotted for the main-sequence galaxies \citep[e.g.][]{Riechers10,Daddi15,Bolatto15,Brisbin19,Brocal21,Boogaard20} and the median value for the SMGs \citep{Sharon16}. Interestingly, we find evidence that all galaxies at $z > 2$ show higher excitation of the $J=3$ level than main-sequence galaxies at $z \approx 1.5$, by a factor of $\approx 1.5-2$. This is consistent with the result from the ALMA Spectroscopic Survey in the Hubble Ultra Deep Field (ASPECS), that galaxies at $z \geq 2$ have an intrinsically higher CO excitation than those at $z < 2$ \citep{Boogaard20}. As noted by the latter authors, the likely cause of the higher CO excitation is the higher SFR surface density $\Sigma_{\rm SFR}$ in higher-$z$ galaxies \citep[e.g.][]{Shibuya15}: simulations have found that the CO excitation is closely linked to the SFR surface density, with a higher excitation obtained for higher values of $\Sigma_{\rm SFR}$ \citep{Narayanan14,Bournaud15}. Fig.~\ref{fig:co31}[B] plots r$_{31}$ versus SFR surface density for our \hi-selected galaxies and the main-sequence galaxies\footnote{The $\Sigma_{\rm SFR}$ values for the main-sequence galaxies are from \citet{Tacconi13} and \citet{Boogaard20}.} of the left panel. It is clear that the SFR surface densities of the \hi-selected galaxies and the $z > 2$ main-sequence galaxies are higher by factors of $\approx 2-10$ than those of the main-sequence galaxies at $z \approx 1.5$, consistent with the higher CO excitation of the higher-$z$ sample.

\subsection{Stellar properties of DLA~B1228-113}

Finally, our SFR estimate of $4.44 \pm 0.47 \ \Msun$~yr$^{-1}$ for the $z \approx 2.1933$ \hi-selected galaxy DLA~B1228-113 from its rest-frame NUV continuum is consistent with the estimate of $\approx 3.9\ \Msun$~yr$^{-1}$ from the H$\alpha$ line \citep{Neeleman18}. The total SFR, estimated from both the total infrared luminosity and fits to the broadband SED, is far higher than the above estimates, $\approx (87-100) \ \Msun$~yr$^{-1}$ \citep{Neeleman18,Klitsch22}, implying a high dust extinction.  Fig.~\ref{fig:images}[A] shows that the rest-frame NUV emission detected in the HST WFC3 image is  offset from the peak of the ALMA CO(3--2) emission, by $\approx 8$~kpc. Combined with the very large CO(3--2) line FWHM \citep[$\approx 600$~\kmps; ][]{Neeleman18}, this suggests that the CO emission may arise from two merging galaxies, one of which has a high extinction and is hence not visible in the HST WFC3 image. High angular resolution CO mapping studies would be of much interest to directly probe this issue. Finally, we cannot formally rule out the possibility that the detected NUV emission arises from an interloper at a different redshift; if so, this would imply an even higher dust extinction for the $z \approx 2.1933$ \hi-selected galaxy.

\section{Summary}

We report JVLA detections of CO(1--0) emission in two \hi-selected galaxies at $z \approx 2.1933$ and $z \approx 2.5848$, and an upper limit to the CO(1--0) line luminosity in a third \hi-selected galaxy at $z \approx 1.9615$. These are the first detections of CO(1--0) emission in \hi-selected galaxies at high redshifts, $z \gtrsim 2$, allowing us to directly estimate the molecular gas mass of the galaxies without assumptions about their CO excitation. We obtain molecular gas masses $\approx 1.5-2$ times lower than earlier estimates based on the mid-$J$ CO lines. We find thermal excitation of the $J=3, 4$ rotational levels in the $z = 2.5848$ galaxy DLA~J0918+1636, and near-thermal excitation of the $J=3$ level in the $z = 2.1933$ galaxy DLA~B1228-113. In both cases, the CO excitation of the mid-$J$ rotational levels is higher than that typically assumed for main-sequence galaxies at $z \approx 2$. We also find evidence for higher excitation of the $J=3$ level in \hi-selected galaxies, main-sequence galaxies, and SMGs at $z \gtrsim 2$, than in main-sequence galaxies at $z \approx 1.5$. This appears to arise due to a higher SFR surface density in the former types of galaxies. Finally, we used HST WFC3 imaging to detect rest-frame NUV emission from the $z \approx 2.1933$ galaxy DLA~B1228-113, obtaining an NUV SFR of $4.44 \pm 0.47 \ \Msun$~yr$^{-1}$. This is a factor of $\approx 20$ lower than the SFR estimated from the total infrared luminosity or from SED fitting, confirming that the object is a highly dusty galaxy. The stellar NUV emission appears slightly offset from the ALMA CO(3--2) emission, suggesting that the CO and NUV emission may arise from a pair of merging galaxies, one of which is highly obscured and hence not detected in its stellar emission.

\software{DrizzlePac \citep{hack20}, {\sc casa} \citep[v5.6;][]{CASA}, {\sc aips} \citep{Greisen03}, {\sc astropy} \citep{astropy13}.}

\begin{acknowledgements}
We thank Johan Fynbo for providing us with the HST fit results for DLA~J0918+1636. BK and NK acknowledge the Department of Atomic Energy for funding support, under project 12-R\&D-TFR-5.02-0700. MN acknowledges support from ERC advanced grant 740246 (Cosmic\_Gas). This research is based on observations made with the NASA/ESA Hubble Space Telescope obtained from the Space Telescope Science Institute, which is operated by the Association of Universities for Research in Astronomy, Inc., under NASA contract NAS 5–26555. Support for Program number 15882 was provided through a grant from the STScI under NASA contract NAS5-26555. The National Radio Astronomy Observatory is a facility of the National Science Foundation operated under cooperative agreement by Associated Universities, Inc. This material is based upon work supported by the National Science Foundation under Grant No. 2107989.
\end{acknowledgements}

\bibliography{CO}{}
\bibliographystyle{aasjournal}

\end{document}